\documentclass{anstrans}
\usepackage{microtype}
\usepackage[utf8]{inputenc}
\usepackage{amsmath}
\usepackage{amsfonts}
\usepackage{amssymb}
\usepackage{graphicx} 
\usepackage{floatrow}
\usepackage[version=4]{mhchem}
\usepackage[dvipsnames]{xcolor}
\usepackage{makecell}
\usepackage[per-mode=symbol,round-precision=3,scientific-notation=false,range-phrase = \text{--}]{siunitx}
\usepackage{booktabs} % nice rules (thick lines) for tables
\usepackage[hyperindex,breaklinks]{hyperref}

\usepackage{floatrow}
\usepackage{xspace}%for abbreviations
\newcommand*{\eg}{e.g.,\@\xspace}
\newcommand*{\ie}{i.e.,\@\xspace}
\newcommand*{\etal}{et al.\@\xspace}
\newcommand*{\vs}{vs.\@\xspace}

\makeatletter
\newcommand*{\etc}{%
    \@ifnextchar{.}%
        {etc}%
        {etc.\@\xspace}%
}
\makeatother

\usepackage[nameinlink,capitalize]{cleveref}
\DeclareUnicodeCharacter{025B}{\ensuremath{\varepsilon}}

\graphicspath{{figures/}}

\usepackage[scaled=0.86]{helvet}
\usepackage[abs]{overpic}
\usepackage{multirow}

\let\mr\mathrm

\DeclareMathOperator*{\argmin}{\arg\!\min}

\title{Numerical Demonstration of Multiple Actuator Constraint Enforcement Algorithm for a Molten Salt Loop}
\author{Akshay J.\@\xspace Dave$^{*}$, Haoyu Wang$^{*}$, Roberto Ponciroli$^{*}$, Richard B.\@\xspace Vilim$^{*}$}

\institute{
$^{*}$Nuclear Science and Engineering, Argonne National Laboratory, Lemont, IL, ajd@anl.gov
}

\begin{document}
%%%%%%%%%%%%%%%%%%%%%%%%%%%%%%%%%%%%%%%%%%%%%%%%%%%%%%%%%%%%%%%%%%%%%%%%%%%%%%%%
\section{Introduction}

To advance the paradigm of autonomous operation for nuclear power plants, a data-driven machine learning approach to control is sought.
Autonomous operation for next-generation reactor designs is anticipated to bolster safety and improve economics.
However, any algorithms that are utilized need to be interpretable, adaptable, and robust.
Interpretable means that one can inspect and understand the underlying relationships between inputs and outputs of the algorithm.
Adaptable refers to the capability of accommodating temporal changes in the underlying system dynamics (\eg due to reactivity swings or fouling).
Robustness refers to stability over long horizons under the presence of sensor or process noise.
In this work, we focus on the specific problem of optimal control during autonomous operation.

\subsection{Objectives}

We will demonstrate an interpretable and adaptable data-driven machine learning approach to autonomous control of a molten salt loop.
To address interpretability, we utilize a data-driven algorithm to identify system dynamics in \textit{state-space representation}.
To address adaptability, a control algorithm will be utilized to modify actuator setpoints while enforcing constant, and \textit{time-dependent} constraints.
Robustness is not addressed in this work, and is part of future work.
To demonstrate the approach, we designed a numerical experiment requiring intervention to enforce constraints during a load-follow type transient.

%%%%%%%%%%%%%%%%%%%%%%%%%%%%%%%%%%%%%%%%%%%%%%%%%%%%%%%%%%%%%%%%%%%%%%%%%%%%%%%%
\section{Methods}

In this section we summarize the key algorithms utilized to demonstrate constraint enforcement.
First, the Reference Governor (RG) algorithm is introduced, alongside a formulation of the dynamics model required to utilize it.
Next, the data-driven algorithm utilized to identify system dynamics, Dynamic Mode Decomposition with Control, is introduced.
Finally, the molten salt loop experiment model used to demonstrate constraint enforcement is presented.

\subsection{Reference Governor}

\begin{figure}
    \includegraphics[width=\linewidth]{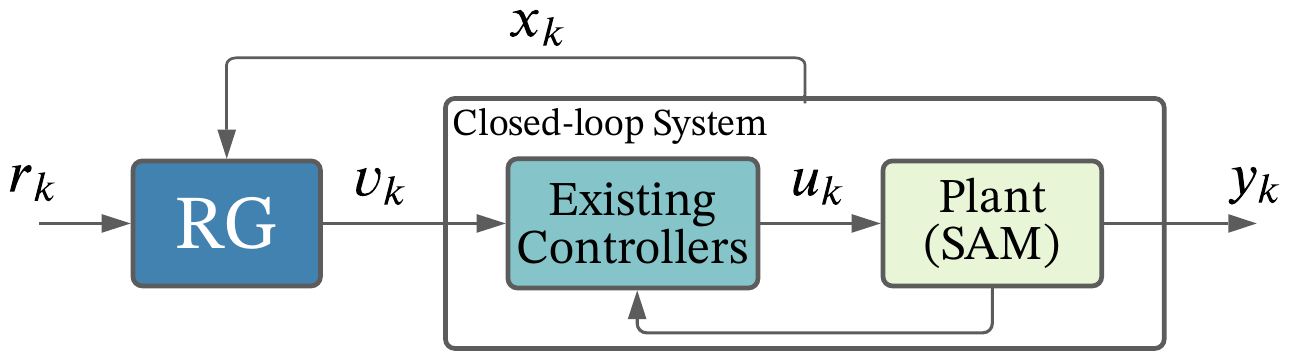}
    \caption{Control block diagram visualizing the topology of the Reference Governor (RG).
    Direct access to the underlying closed-loop system is always available. 
    }
    \label{fig:RG}
\end{figure}

The RG \cite{Bemporad1994} is the constraint enforcement algorithm utilized in this work.
The RG is an \textit{add-on} control scheme that enforces \textit{point-wise in time} constraints.
The terminology \textit{add-on} reflects the nature of its intended use: the algorithm is applied to a system that already has an existing pre-calibrated control logic (\eg regulation via PID) for a closed-loop feedback system.
The add-on feature is important as an operator may choose to manually override/disable the RG, while maintaining controllability to maneuver the plant.
The terminology \textit{point-wise in time} reflects the capability to update constraint bounds at each discrete time-step.
This feature allows us to address \eg degradation phenomena, where the capabilities of the system vary temporally.
The application of RG is visualized in \cref{fig:RG}.
Previous work demonstrated that the RG framework is applicable to an integrated energy system \cite{Wang2020}.

The RG algorithm is summarized next.
Garone \etal \cite{Garone2017} provide a full review.
First, what the RG algorithm does, with respect to \cref{fig:RG} is described.
To gain some intuition, we first describe the entry-level RG: the Scalar Reference Governor (SRG).
At each time-step $k\in\mathbb{Z}_+$, the SRG receives a reference input, $r_k\in\mathbb{R}^m$, where $m$ is the number of inputs.
Depending on the system's current state, expected evolution, and user defined constraints, an admissible input $v_k\in\mathbb{R}^m$,
\begin{equation}
    v_k = v_{k-1} + \kappa_k \left(r_k - v_{k-1}\right)~,
    \label{eq:rg1}
\end{equation}
is sent to the lower-level controllers.
In \cref{eq:rg1}, $\kappa_k\in[0,1]$ is a \textit{scalar} that governs admissible changes to the inputs, such that a complete rejection, $v_k=v_{k-1}$, acceptance, $v_k=r_k$, or an intermediate change, $v_{k-1}\leq v_k\leq r_k$, is possible.
After determining $\kappa_k$, the output of the RG, $v_k$ is sent to the existing controllers.
For example, a PID controller would receive a component of $v_k$ as a reference set-point for its manipulated variable (MV), and according to its paired MV's current value, issue a control action $u_k$ to the control variable (CV).

Next, how the SRG algorithm determines $\kappa_k$ is summarized.
The RG relies on a state-space representation,
\begin{align}
    x_{k+1} & = Ax_k + Bu_k~, \label{eq:state}\\
    y_k & = Cx_k + Du_k~, \label{eq:outp}
\end{align}
where $x_k\in\mathbb{R}^n$, $u_k\in\mathbb{R}^m$, and $y_k\in\mathbb{R}^p$ are vectors at step $k$ representing states, inputs, and outputs, respectively.
The matrix $A\in\mathbb{R}^{n\times n}$ accounts for the state dynamics.
The matrix $B\in\mathbb{R}^{n\times m}$ is the input matrix.
The matrix $C\in\mathbb{R}^{p\times n}$ is the output matrix.
The matrix $D\in \mathbb{R}^{p\times m}$ is the control pass-through matrix.
Using the system model defined in \cref{eq:state,eq:outp}, constraints are imposed on the output variables, $y_k\in Y$, where $Y$ is defined by a set of linear inequalities.
Constraints may be imposed on $x_k$, or $u_k$.
To ensure constraints are not violated, we define the maximal output admissible set (MOAS).
The MOAS is the set of all $x_k$, and constant input $\Tilde{v}$, such that,
\begin{equation}
    O_\infty = {(x_k, \Tilde{v}) : y_{t+k}\in Y, v_{t+k}=\Tilde{v}, \forall k \in \mathbb{Z}^+}~.
    \label{eq:moas}
\end{equation}
The $k\rightarrow \infty$ assumption of $O_\infty$ is generally relaxed to a suitably large finite horizon, $T$.
Thus, at each time-step $k$, the SRG algorithm determines the admissible $\Tilde{v}$, and therefore $\kappa_k$, such that the system remains in $O_\infty$.
The $O_\infty$ is constructed by evaluating \cref{eq:state,eq:outp} for each $k=0,1...T$ time steps.

A drawback of the SRG formulation is that $\kappa_k$ is a scalar.
Therefore, if $v_k-r_k$ is multi-dimensional and $\kappa_k<1$, input movement in all dimensions is bound.
In our work, we utilize the Command Governor (CG), a variant of the RG which selects $v_k$ by solving the quadratic program,
\begin{align}
v_k = &\argmin_{v_k} ||v_k - r_k ||^2_Q~, \nonumber \\
\mr{s.t.} \hspace{6pt} &(x_k, v_k=\Tilde{v})\in O_\infty~, \label{eq:cg}
\end{align}
where $Q$ is a positive definite matrix signifying the relative importance of each $v_k$ component.
In this work, the quadratic program is solved by the CVXPY convex optimization library \cite{diamond2016cvxpy}.
Using the CG, constraints can be enforced by manipulating $v_k$ arbitrarily in $\mathbb{R}^m$, constrained by $O_\infty$.
This is particularly powerful for systems where $m>1$, \ie multiple inputs.

Accurately identifying the system dynamics is a necessary prerequisite for utilizing the RG framework.
In general, and in our particular model, $D=0$.
A system identification method is needed to define the remaining matrices.

\subsection{System Identification}

To identify the system matrices, we utilize the Dynamic Mode Decomposition with Control (DMDc) algorithm \cite{Proctor2016}.
DMDc is an extension of Dynamic Mode Decomposition (DMD), a method that originated in the fluids community to extract spatial-temporal coherent structures from experimental data.
DMDc is a data-driven method relying on snapshots of data, of length $L$,
\begin{equation}
X= \begin{bmatrix}
\vline & \vline & & \vline \\
x_0 & x_1 & \cdots & x_L\\
\vline & \vline & & \vline
\end{bmatrix} ,~~ U = \begin{bmatrix}
\vline & \vline & & \vline \\
u_0 & u_1 & \cdots & u_{L-1} \\
\vline & \vline & & \vline
\end{bmatrix},
\label{eq:snaps}
\end{equation}
corresponding to the states and inputs defined in \cref{eq:state}.
The data in \cref{eq:snaps} is grouped into two matrices: $\Omega=[X_{0...L-1}^\intercal U^\intercal]^\intercal$ and $X'=X_{1...L}$.
Using the singular value decomposition (SVD) of $\Omega$ and $X'$, the reduced order approximations of matrices $A$ and $B$ can be determined  \cite[\S3.3]{Proctor2016}.
Qualitatively, the DMDc methodology combines SVD (dimensionality reduction) with a Fourier transform (finding dominant patterns in time).

\subsection{Experiment Model}

\begin{figure}
        \begin{minipage}[t]{0.5\linewidth}
        \vspace{0pt}
        \centering
        \includegraphics[trim={0 0 0 0.1cm}, clip, width=0.9\linewidth]{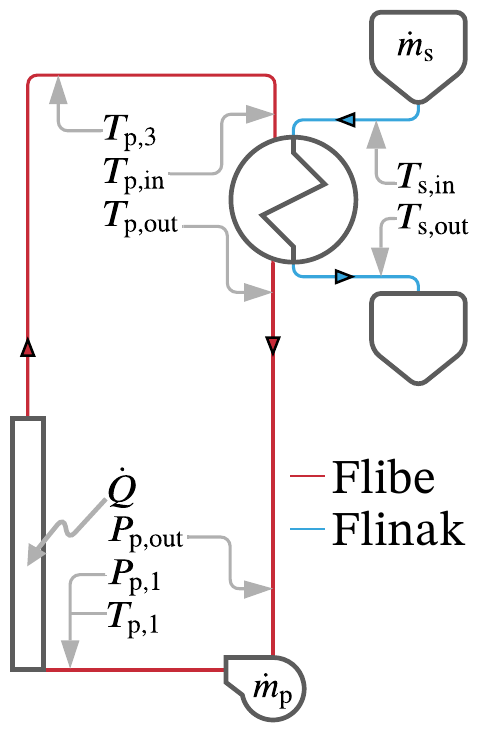}
        \end{minipage}%
        \begin{minipage}[t]{0.5\linewidth}
        \vspace{0pt}
        \centering
         \begin{tabular}{lc}
             \hline
             \multicolumn{2}{c}{Steady-state Value} \\
             \hline
             \multicolumn{2}{l}{Primary (flibe)} \\
              \hspace{6pt}$\dot{Q}$ [MW] & 18.0 \\
              \hspace{6pt}$\dot{m}_\mr{p}$ [kg/s] & 589 \\
              \hspace{6pt}$T_\mr{p,in}$ [\SI{}{\celsius}] & 585 \\
              \hspace{6pt}$T_\mr{p,out}$ [\SI{}{\celsius}] & 572 \\
              \hspace{6pt}$T_\mr{p,1}$ [\SI{}{\celsius}] &  572 \\
              \hspace{6pt}$T_\mr{p,3}$ [\SI{}{\celsius}] &  585 \\
              \hspace{6pt}$P_\mr{p,out}$ [kPa] & 179 \\
              \hspace{6pt}$P_\mr{p,1}$ [kPa] & 200 \\
            \multicolumn{2}{l}{Secondary (flinak)} \\
              \hspace{6pt}$\dot{m}_\mr{s}$ [kg/s] & 380 \\
              \hspace{6pt}$T_\mr{s,in}$ [\SI{}{\celsius}] & 492 \\
              \hspace{6pt}$T_\mr{s,out}$ [\SI{}{\celsius}] & 517 \\
         \end{tabular}
        \end{minipage}
    \caption{%
        Overview of the experiment model.
        Left: Layout of the model in SAM.
        Right: Corresponding steady-state conditions.
    }
    \label{fig:loop}
\end{figure}

This work focuses on demonstrating the applicability of advanced control algorithms for next-generation reactors.
Thus, the System Analysis Module (SAM) code \cite{Hu2017} is utilized.
SAM is developed to provide best-estimate system-level models for Sodium-cooled Fast Reactors, Molten Salt Reactors, Fluoride-cooled High-temperature Reactors, and Lead-cooled Fast Reactors.
A layout of the experiment model and its steady-state characteristics is presented in \cref{fig:loop}.
The model consists of a single loop filled with flibe (\ce{2LiF-BeF_2}), heated through a single channel and cooled with flinak (\ce{LiF-NaF-KF}) via a heat exchanger.
Thermophysical properties of flibe and flinak are embedded within SAM.
In this experiment, there are two fixed boundary conditions: the flinak source temperature ($T_\mr{s,in}$) and the primary-side pump head.

% Experiment model and SS values

After the model was defined within SAM, a transient to demonstrate the capabilities of the RG was formulated.
A load-follow type transient was chosen.
A pre-defined reference trajectory is assigned for the secondary side mass flow rate ($\dot{m}_\mr{s,ref}$) and the primary-side heat-exchanger inlet temperature ($T_\mr{p,in,ref}$).
A control strategy for the maneuver was then developed, alongside categorization of parameters to identify the system matrices.

\subsection{Control Strategy \& State Selection}

The control strategy encompasses the pairing of CVs to MVs.
Effective pairings minimize cross-talk, which can be quantified via open loop response.
Remaining (uncontrolled) variables are to be categorized into outputs and states.
All outputs are variables that can be directly measured from the plant and that may be subject to constraints.
The states are variables that may not be directly measured from the plant (\eg the integral component of the PI controller for secondary-side mass-flow rate, $I_{\dot{m}_\mr{s}}^\mr{PI}$). 
The composition of the set of states is determined using a feature-selection algorithm.
A forward sequential feature selection algorithm was utilized.
The objective function to minimize was the mean squared error between the output values from the state-space model, \ie \cref{eq:state,eq:outp}, and the SAM model.
The resulting control strategy and classification of variables are summarized in \cref{tab:vars}.
The RG reference inputs, $r_k$ in \cref{eq:rg1}, are defined as the actuators of the entire system.

% Control strategy table
\begin{table}%
    \begin{tabular}{ccc}%
        \begin{tabular}[t]{l}
            \hline
            Outputs -- $y_k$\\
            \hline
             $T_\mr{p,out}$ \\
             $T_\mr{s,out}$ \\
             $P_\mr{p,out}$ \\
             $P_\mr{p,1}$ \\
        \end{tabular} &
        \begin{tabular}[t]{l}
            \hline
            States -- $x_k$\\
            \hline
             $T_\mr{p,1}$ \\
             $T_\mr{p,3}$ \\
             $I_{\dot{m}_\mr{s}}^\mr{PI}$
        \end{tabular} &
        \begin{tabular}[t]{l}
            \hline
            Inputs -- $r_k$\\
            \hline
            $\dot{m}_\mr{s,ref}$ \\
            $T_\mr{p,in,ref}$
        \end{tabular}
    \end{tabular}
    \begin{tabular}{ll}
        \begin{tabular}[t]{ll}
            \hline
            CVs -- $u_k$ & MVs -- $v_k$\\
            \hline
            $u_\mr{s}$ & $\dot{m}_\mr{s}$ \\
            $\dot{Q}$ & $T_\mr{p,in}$ 
        \end{tabular} & 
        \begin{tabular}[t]{l}
            \hline
            Constrained Variables\\
            \hline
            $T_\mr{p,out}\leq T_\mr{p,out}^\mr{max}$\\
            $T_\mr{s,out}\geq T_\mr{s,out}^\mr{min}$
        \end{tabular}
    \end{tabular}
    \caption{%
    Categorization of variables corresponding to the labels in \cref{fig:RG} and model parameters in \cref{fig:loop}.
    First major row presents the list of outputs, states and inputs.
    Second major row presents the control variables (CV), manipulated variables (MV), and constrained variables.
    }
    \label{tab:vars}
\end{table}

%%%%%%%%%%%%%%%%%%%%%%%%%%%%%%%%%%%%%%%%%%%%%%%%%%%%%%%%%%%%%%%%%%%%%%%%%%%%%%%%
\section{Results}

\raggedbottom
The model implemented in SAM, \cref{fig:loop}, and the control architecture, \cref{fig:RG}, were coupled via python.
A Proportional-Integral (PI) controller was constituted for both MVs tabulated in \cref{tab:vars}, with the CVs as outputs.
For all results presented in this paper, the topology of \cref{fig:RG} is maintained where `Plant' is assumed to be the ground-truth results from SAM and `Existing Controllers' are the PI controllers that were tuned via open-loop response. 
A constant $\Delta t \equiv \left(t_{k+1}-t_k\right) = \SI{0.2}{\second}$ was maintained.

\subsection{SAM \vs DMDc}

\begin{figure}
    \centering
    \includegraphics[width=\linewidth]{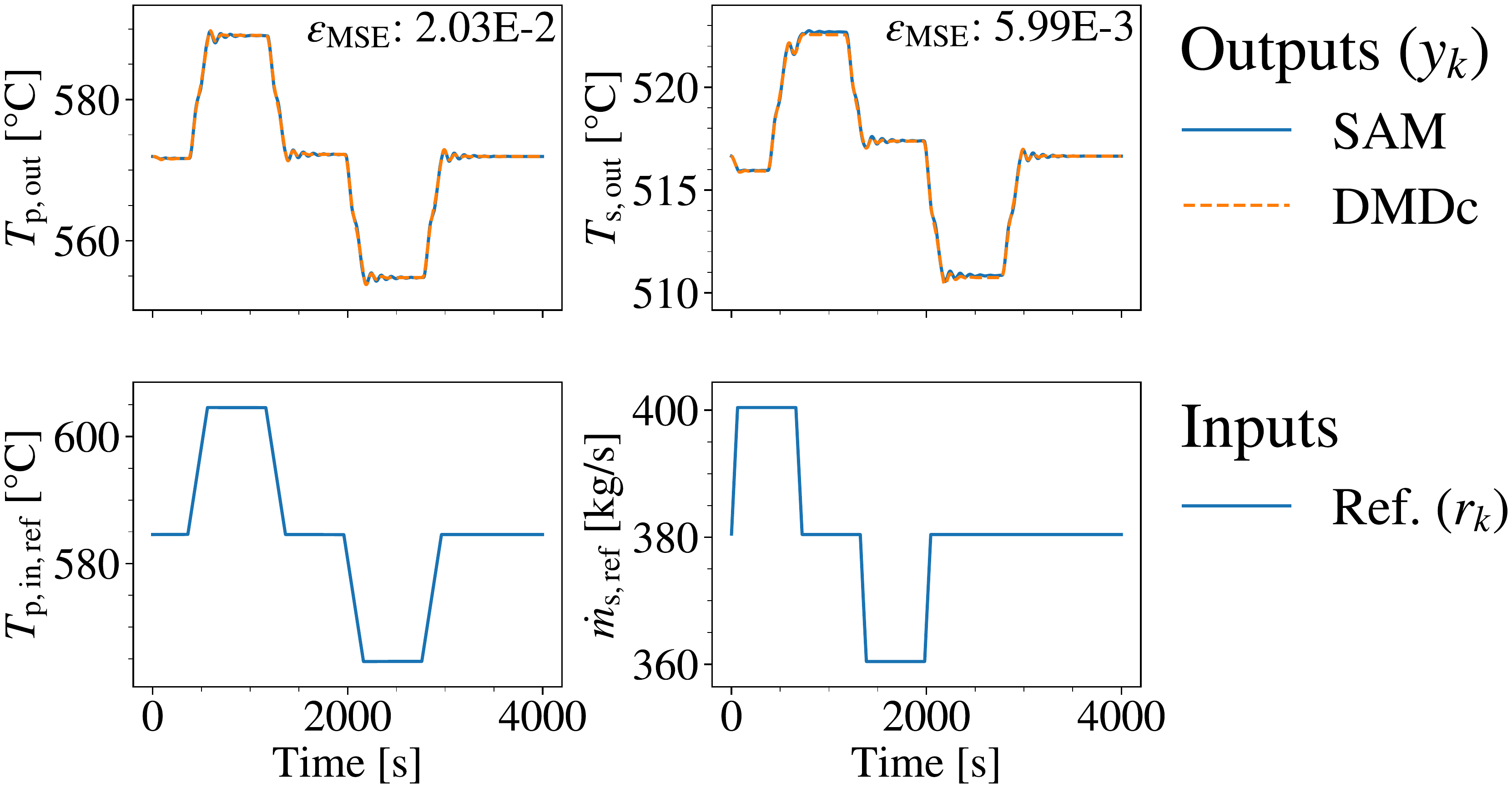}
    \caption{Application of DMDc to identify state-space matrices.
    Top rows: Contrasting DMDc model performance \vs SAM.
    Mean-squared error ($\varepsilon_\mr{MSE}$) is presented.
    Bottom row: Reference setpoint trajectory -- the input for both SAM and DMDc.
    }
    \label{fig:res_dmdc}
\end{figure}

First, the validity of applying DMDc to identify the SAM model dynamics is discussed.
To simulate a load-follow transient, time-dependent reference trajectories were assigned to $T_\mr{p,out,ref}$ and $\dot{m}_\mr{s,ref}$.
The trajectories are specified such that $T_\mr{p,out}^\mr{max}\equiv \SI{586.85}{\celsius}$ and $T_\mr{s,out}^\mr{min}\equiv \SI{512.85}{\celsius}$ constraints are violated.
These trajectories are presented in the bottom row of \cref{fig:res_dmdc}.
Once the input ($u_k$), state ($x_k$) and output ($y_k$) data is collected, we utilize the DMDc algorithm to identify the matrices in the state-space representation, \cref{eq:state,eq:outp}.
A comparison of the state-space representation to results from SAM are presented in the top row of \cref{fig:res_dmdc}.
By visual inspection, it is apparent that the DMDc algorithm successfully identifies the system dynamics.
Quantitatively, the mean-squared error loss for each state and output is also low.
On additional tests with alternating ramp sequences our algorithms achieved equivalent performance -- we present results for a single sequence for brevity.

\subsection{Constant Constraint Enforcement}

\begin{figure}
    \centering
    \includegraphics[width=\linewidth]{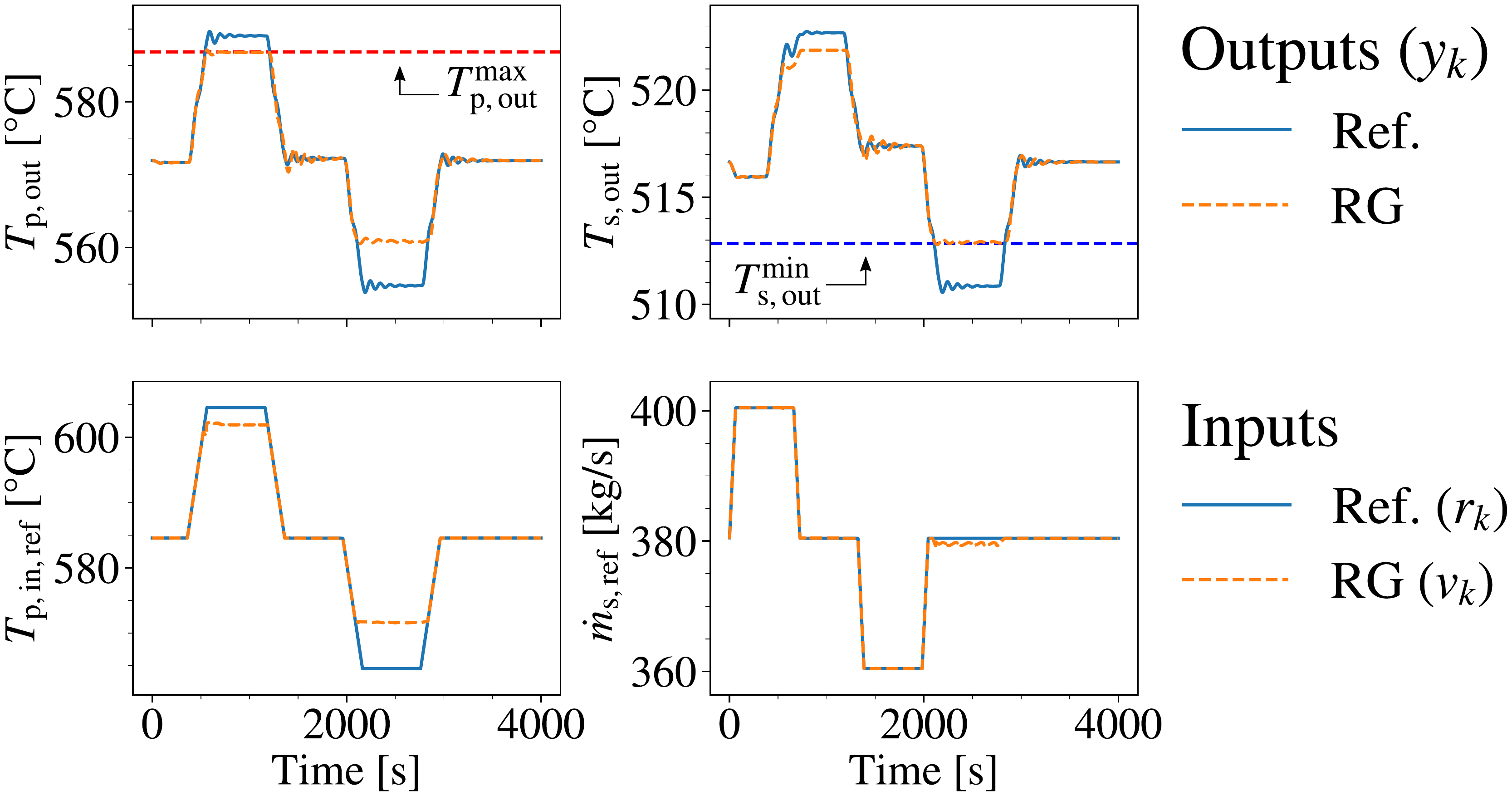}
    \caption{Results with application of RG algorithm with constant constraint values.
    Top row displays outputs that are subject to constraints.
    Bottom row displays changes to the reference input trajectory by the RG algorithm.
    }
    \label{fig:cg}
\end{figure}

The first result presented utilizes constant constraint values.
Because the constraint values are constant, the admissible region changes only with respect to $x_k$.
The results are presented in \cref{fig:cg}.
The results indicate that the RG algorithm correctly intervenes and restricts $v_k$ such that the constraints assigned to both outputs are enforced.
It is important to note that the results presented in the following figures are from manipulating a SAM plant, rather than the DMDc model.

\subsection{Time-dependent Constraint Enforcement}

\begin{figure}
    \centering
    \includegraphics[width=\linewidth]{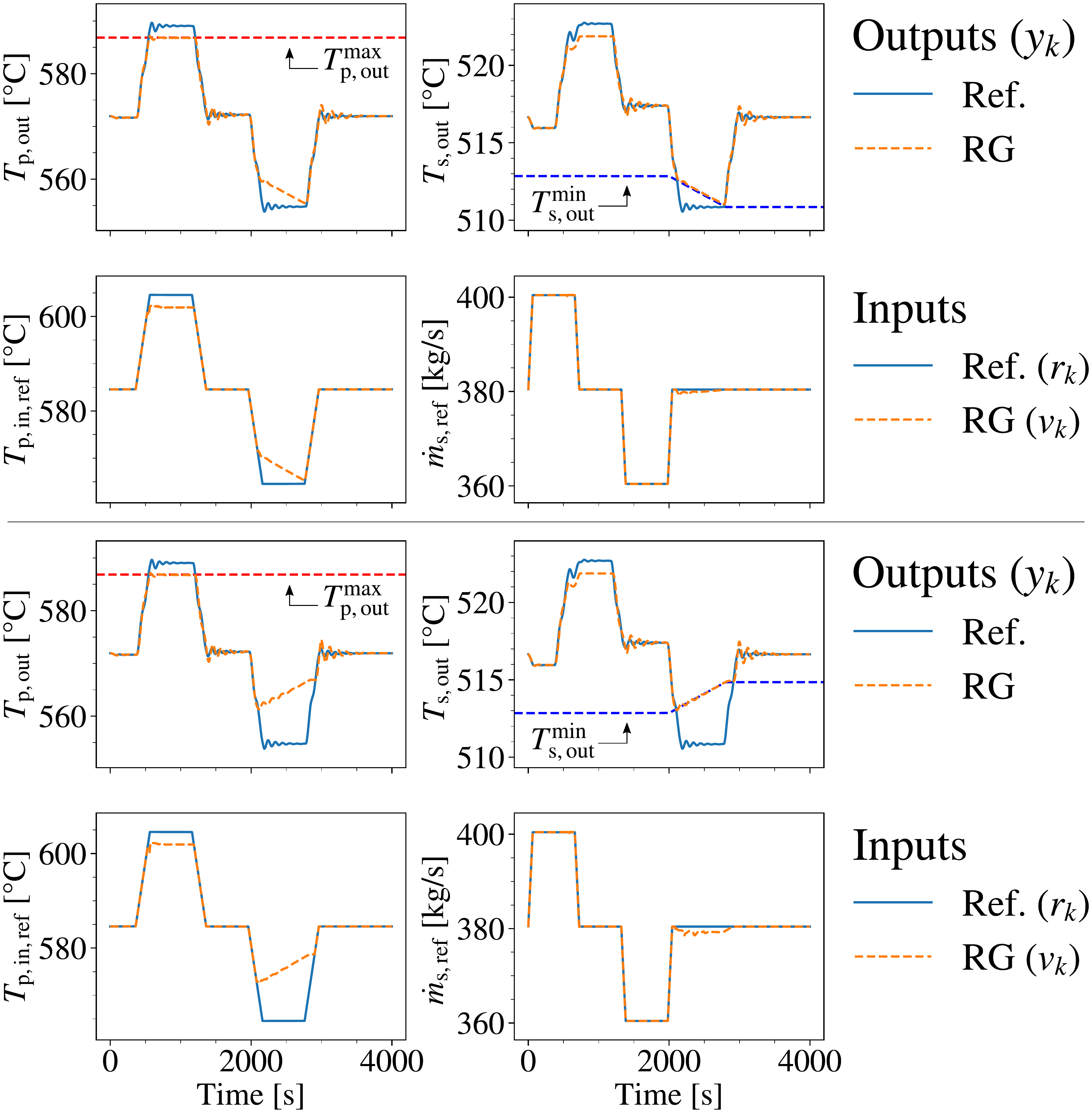}
    \caption{Results with application of RG algorithm with constant $T_\mr{p,out}^\mr{max}$, decreasing $T_\mr{s,out}^\mr{min}$ (top two rows) and increasing $T_\mr{s,out}^\mr{min}$ (bottom two rows).
    }
    \label{fig:cg_tdep}
\end{figure}

The next set of results utilize time-dependent constraints.
In particular the secondary-side outlet temperature is constrained such that,
\begin{equation}
  T_\mr{s,out}^\mr{min}=\begin{cases}
       512.85~\SI{}{\celsius}, &t< \SI{2000}{\second}\\
       512.85\pm\frac{2.5}{1000}(t-2000)~\SI{}{\celsius}, &\SI{2000}{\second}\leq t\leq \SI{2800}{\second}\\
       512.85\pm2.00~\SI{}{\celsius}, &t> \SI{2800}{\second}
    \end{cases}
    \label{eq:constr}
\end{equation}
where the $+~\mr{or}~-$ sign is used consistently for increasing or decreasing $T_\mr{s,out}^\mr{min}$, respectively.
Results for decreasing or increasing the constraint is presented in \cref{fig:cg_tdep}.
First, the scenario where $T_\mr{s,out}^\mr{min}$ decreases is discussed.
As the constraint decreases, the admissible region ($O_\infty$) expands as the admissible setpoint for $T_\mr{p,in,ref}$ can be reduced further to meet the reference value. 
The results in \cref{fig:cg_tdep} indicate that the RG algorithm accurately exploits the expansion of $O_\infty$ to minimize the cost in \cref{eq:cg}.
The more challenging scenario, where $T_\mr{s,out}^\mr{min}$ increases is discussed next.
During this scenario, the $O_\infty$ region contracts as the admissible setpoint for $T_\mr{p,in,ref}$ increases in distance from the reference value.
The scenario is more challenging as an incorrect $v_{k+1}$ can immediately violate constraints.
The results in \cref{fig:cg_tdep} indicate that the RG algorithm also optimally responds to the contraction of $O_\infty$.

\subsection{Admissible Region Evolution}

\begin{figure}
    \centering
    \includegraphics[width=\linewidth]{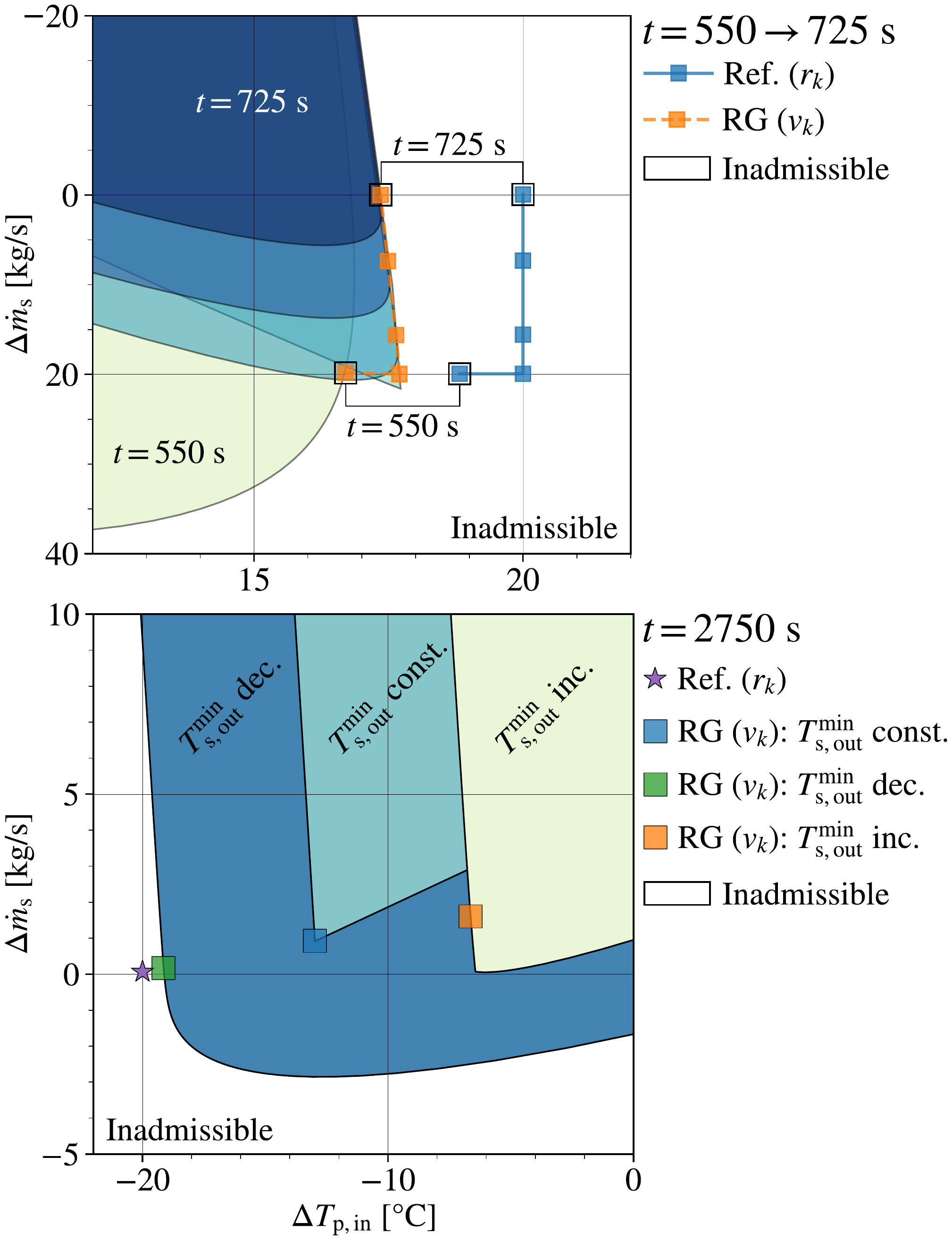}
    \caption{Visualizations of $O_\infty$.
    Top: $O_\infty\left( t=550\rightarrow \SI{725}{\second}\right)$ with constant constraints.
    Bottom: Contrast of $O_\infty\left(t=\SI{2750}{\second}\right)$ with constant (const.), decreased (dec.), and increased (inc.) values of $T_\mr{s,out}^\mr{min}$, according to \cref{eq:constr}.
    In both figures, the axes represent input values relative to steady-state, \cref{fig:loop}.
    } 
    \label{fig:moas}
\end{figure}

The admissible region ($O_\infty$) evolution is discussed next.
In \cref{fig:moas}, two separate events are presented.
The evolution of $O_\infty\left( t=550\rightarrow \SI{725}{\second}\right)$ visualizes the geometric objective of the CG, \cref{eq:cg}, a minimization of the distance between $v_k$ and $r_k$ as the bounds of $O_\infty$ evolve.
Note that the evolution of $O_\infty$ in this case occurs solely due to the change in $x_k$, as constraints are constant.
Whereas, the contrast of $O_\infty\left(t=\SI{2750}{\second}\right)$ for the different trajectories of $T_\mr{s,out}^\mr{min}$ visualize the expansion and contraction of $O_\infty$ referred to previously.
In this case, the differences in $O_\infty$ is due to different $x_k$ \textit{and} $T_\mr{s,out}^\mr{min}$ values.

%%%%%%%%%%%%%%%%%%%%%%%%%%%%%%%%%%%%%%%%%%%%%%%%%%%%%%%%%%%%%%%%%%%%%%%%%%%%%%%%
\section{Concluding Remarks}

To advance the paradigm of autonomous operation for nuclear power plants, a data-driven machine learning approach to control is sought.
This work demonstrates an interpretable and adaptable approach to constraint enforcement for a molten salt loop, with multiple actuators and time-dependent constraints.
The major contribution of this works are: developing a framework to interface DMDc and RG algorithms with SAM; demonstrating that DMDc provides accurate state-space representations for load-follow type transients for flibe-flinak loops; and demonstrating successful actuator setpoint intervention to enforce constant and time-dependent constraints using the RG framework.
The next steps of our work include: introduction of noise to gauge robustness of the proposed framework, and application to a SAM plant with nuclear dynamics.

%%%%%%%%%%%%%%%%%%%%%%%%%%%%%%%%%%%%%%%%%%%%%%%%%%%%%%%%%%%%%%%%%%%%%%%%%%%%%%%%
\section{Acknowledgments}
This work is supported by U.S. Department of Energy Nuclear Energy Enabling Technologies Project 20-19321, and 
U.S. Department of Energy Advanced Research Projects Agency–Energy Project 2174-1556.

%%%%%%%%%%%%%%%%%%%%%%%%%%%%%%%%%%%%%%%%%%%%%%%%%%%%%%%%%%%%%%%%%%%%%%%%%%%%%%%%
\bibliographystyle{ans}
\bibliography{ANS2022_DMDc_CG.bib}

\end{document}